\documentclass[prd,preprint,superscriptaddress,eqsecnum,showpacs,nofootinbib,nobibnotes,noeprint]{revtex4-1}
\usepackage{amsfonts,amssymb,amsmath,dsfont,epsfig}
\usepackage{amsfonts,amssymb,amsmath}
\usepackage{graphicx}
\usepackage{paralist}
\usepackage[english]{babel}
\usepackage{slashed}
\usepackage[usenames]{xcolor}
\usepackage{xspace}

\usepackage[linkcolor=blue,citecolor=blue,urlcolor=blue,colorlinks=true]{hyperref} 

\usepackage{etoolbox}

\makeatletter
\patchcmd{\frontmatter@abstract@produce}
  {\vskip200\p@\@plus1fil
   \penalty-200\relax
   \vskip-200\p@\@plus-1fil}
  {}
  {}
  {}
\makeatother

\def\ie{{\it i.e.}, }
\def\eg{{\it e.g.}, }

\newcommand{\be}{\begin{equation}}
\newcommand{\ee}{\end{equation}}

\def\1eq#1{Eq.~(\ref{#1})}

\def\2eqs#1#2{Eqs.~(\ref{#1}) and~(\ref{#2})}
\def\3eqs#1#2#3{Eqs.~(\ref{#1}),~(\ref{#2}) and~(\ref{#3})}

\def\fig#1{Fig.~\ref{#1}}

\newcommand{\Gnp}{\Gamma}
\newcommand{\Ls}{ \mathit{L}_{{sg}}}      

\begin{document}

\title{Ghost dynamics from Schwinger-Dyson equations}

\author{M.~N. Ferreira}
\affiliation{\mbox{University of Campinas - UNICAMP, Institute of Physics ``Gleb Wataghin'',} 13083-859 Campinas, S\~{a}o Paulo, Brazil.}


\begin{abstract}

We discuss the coupled dynamics of the ghost dressing function and the ghost-gluon vertex through the Schwinger-Dyson equations that they satisfy. In order to close the system of equations, we combine recent lattice data for the gluon propagator and an approximate STI-derived Ansatz for the general kinematics three-gluon vertex. The numerical solution of the resulting coupled system exhibits excellent agreement to lattice data, for both the ghost dressing function and the ghost-gluon vertex, and allows the determination of the coupling constant. Next, in the soft gluon limit the full three-gluon vertex appearing in the ghost-gluon equation reduces to a special projection that is tightly constrained by lattice simulations. Specializing the ghost-gluon Schwinger-Dyson equation to this limit provides a nontrivial consistency check on the approximations employed for the three-gluon interaction and shows that the latter has an important quantitative effect on the ghost-gluon vertex. Finally, our results stress the importance of eliminating artifacts when confronting lattice data with continuum predictions.

\end{abstract}

\maketitle


\section{Introduction}

In the context of continuum methods for the study of nonperturbative QCD~\cite{Roberts:1994dr,Alkofer:2000wg,Maris:2003vk,Pawlowski:2005xe,Fischer:2006ub,Aguilar:2006gr,Aguilar:2008xm,Boucaud:2008ky,Binosi:2009qm,Campagnari:2010wc,Pennington:2011xs,Vandersickel:2012tz,Serreau:2012cg,Cloet:2013jya,Kondo:2014sta,Binosi:2016rxz,Cyrol:2017ewj,Corell:2018yil,Huber:2018ned,Aguilar:2019jsj,Pelaez:2021tpq,Eichmann:2021zuv}, such as Schwinger-Dyson equations (SDEs), the ghost sector plays a distinguished role for the relative simplicity of its dynamical equations. In particular, the ghost sector SDEs involve a reduced number of other Green's functions and, in Landau gauge, benefit from the Taylor theorem~\cite{Taylor:1971ff}, which facilitates their nonperturbative renormalization~\cite{Boucaud:2008ky,Boucaud:2008gn,Blossier:2010ky,Aguilar:2013xqa}. As such, they furnish ideal testing grounds to assess the reliability of truncation schemes, probe the impact of other Green's functions, and evaluate the consistency of SDEs with other methods, such as lattice simulations.

Furthermore, understanding the infrared (IR) behavior of the ghost propagator and vertices themselves is important due to their connections to proposed scenarios of color confinement~\cite{Kugo:1979gm,Nakanishi:1990qm} and because they affect other Green's functions~\cite{Cucchieri:2008qm,Alkofer:2008jy,
Huber:2012zj,Aguilar:2013vaa,Blum:2014gna,Eichmann:2014xya,Blum:2015lsa,Cyrol:2016tym,Duarte:2016ieu,Athenodorou:2016oyh,Boucaud:2017obn,Aguilar:2019jsj,Aguilar:2019uob,Aguilar:2019kxz,Aguilar:2021lke}. In particular, the nonperturbative masslessness of the ghost~\cite{Aguilar:2008xm,Boucaud:2008ky} implies the vanishing of the gluon spectral function at the origin~\cite{Cyrol:2018xeq,Horak:2021pfr} and causes IR divergences in the three-gluon vertex~\cite{Aguilar:2013vaa,Aguilar:2019jsj}, which contribute to the observed suppression of this function at small energies~\cite{Huber:2012zj,Aguilar:2013vaa,Eichmann:2014xya,Blum:2015lsa,Athenodorou:2016oyh,Duarte:2016ieu,Boucaud:2017obn,Corell:2018yil,Aguilar:2019jsj,Aguilar:2019uob,Aguilar:2019kxz,Huber:2020keu,Aguilar:2021lke,Catumba:2021yly}.

In the present work, we solve the coupled system of SDEs governing the momentum evolution of the ghost propagator, $D(p^2)$, and the form factor, denoted by $B_1(r,p,q)$, of the classical (tree level) tensor structure of the ghost-gluon vertex. For the gluon propagator which appears as ingredient, we capitalize on lattice results~\cite{Bogolubsky:2009dc,Boucaud:2018xup,Aguilar:2021okw}, carefully extrapolated to the continuum and to infinite volume~\cite{Boucaud:2018xup,Aguilar:2021okw} and displaying the now firmly established IR saturation~\cite{Cucchieri:2007rg,Bogolubsky:2009dc,Sternbeck:2012mf,Boucaud:2011ug,Oliveira:2010xc,Ayala:2012pb,Bicudo:2015rma,Aguilar:2019uob}, associated with the dynamical generation of a gluon mass gap~\cite{Cornwall:1981zr,Aguilar:2006gr,Aguilar:2008xm,Aguilar:2019kxz,Aguilar:2020uqw}. In this way, we are left with the three-gluon vertex, $\Gamma_{\alpha\mu\nu}(q,r,p)$, as the most uncertain ingredient, whose nonperturbative structure has only recently begun to be unraveled~\cite{Huber:2012zj,Aguilar:2013vaa,Eichmann:2014xya,Blum:2015lsa,Athenodorou:2016oyh,Duarte:2016ieu,Boucaud:2017obn,Corell:2018yil,Aguilar:2019jsj,Aguilar:2019uob,Aguilar:2019kxz,Huber:2020keu,Aguilar:2021lke,Catumba:2021yly}.

Then, we use our system of equations to indirectly probe the impact of $\Gamma_{\alpha\mu\nu}(q,r,p)$. To this end, we employ two different methods: first, we implement an approximation derived from the Slavnov-Taylor identity (STI) that $\Gamma_{\alpha\mu\nu}(q,r,p)$ satisfies~\cite{Ball:1980ax,Aguilar:2019jsj} and which captures its main known features; next, in the soft gluon limit the entire contribution of $\Gamma_{\alpha\mu\nu}(q,r,p)$ to our system of SDEs reduces~\cite{Aguilar:2021okw} to a special projection, denoted by $\Ls(q^2)$, which has been accurately determined on the lattice~\cite{Athenodorou:2016oyh,Boucaud:2017obn,Aguilar:2021lke}. Comparing the two solutions in the soft gluon limit, we find nearly prefect agreement, thus validating the approximation employed for the general kinematics $\Gamma_{\alpha\mu\nu}(q,r,p)$. 

With the approximation for the input three-gluon vertex validated in the above way, we show that $\Gamma_{\alpha\mu\nu}(q,r,p)$ has an important quantitative impact on $B_1(q,r,p)$. Moreover, the SDE results for both the ghost propagator and ghost-gluon vertex are found to agree strikingly with lattice data.

\section{Coupled system of SDEs}

\begin{figure}[h]
  \includegraphics[width=0.5\linewidth]{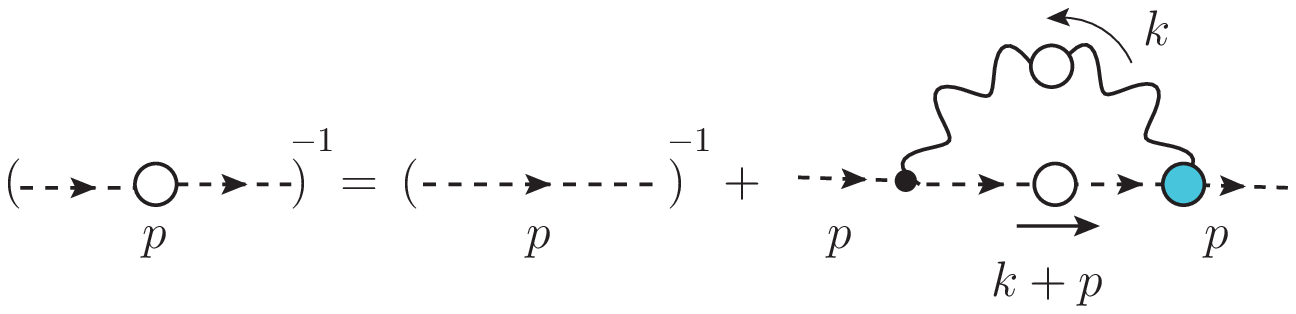}\\
  \vspace{0.2cm}
  \includegraphics[width=0.75\linewidth]{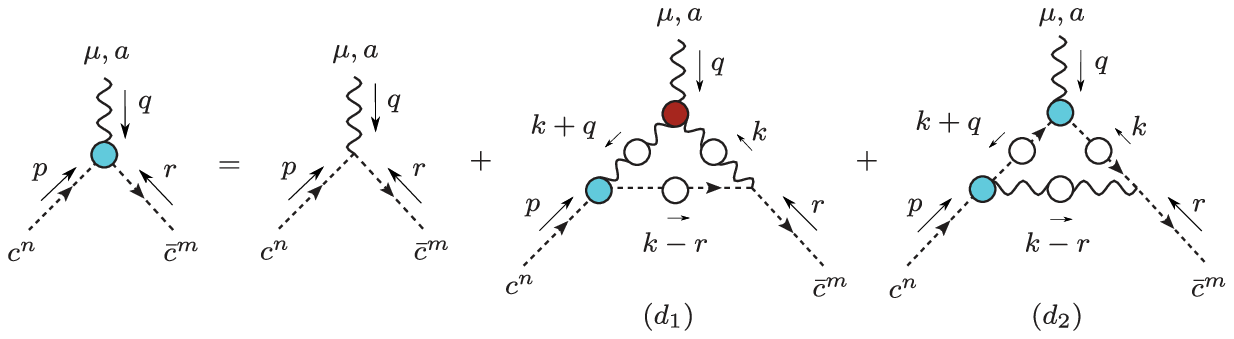}
  \caption{Coupled system of SDEs for the ghost propagator (top) and ghost-gluon vertex (bottom). Wavy and dashed lines represent gluon and ghost fields, respectively, white circles represent fully dressed propagators, whereas blue and red circles correspond to dressed ghost-gluon and three-gluon vertices, respectively.\label{cpld}}
\end{figure}

The coupled system of SDEs which will be the focal point of this study is shown diagrammatically in \fig{cpld}. Note that, while the SDE for the ghost propagator (top line) is left intact, the equation for the ghost-gluon vertex is truncated at the ``one loop dressed'' level, where we neglect one diagram containing a 4-point function, which has been been shown to have only a $2\%$ effect on the outcome of this SDE~\cite{Huber:2017txg,Huber:2018ned}.

It is convenient to factor out the tree level form from the ghost propagator, $D^{ab}(p^2) = i \delta^{ab}D(p^2)$, to define the ghost dressing function, $F(p^2)$, through \mbox{$D(p^2) = F(p^2)/p^2$}. For the ghost-gluon vertex, $\Gnp_\mu^{abc}(r,p,q) = -gf^{abc}\Gnp_\mu(r,p,q)$, where $r$, $p$ and $q$ denote the anti-ghost, ghost, and gluon momenta, respectively, we decompose $\Gnp_\mu(r,p,q)$ into its most general tensor form
\be 
\Gnp_\mu(r,p,q) = r_\mu B_1(r,p,q) + q_\mu B_2(r,p,q) \,.
\ee
Note that at tree level, $B_1^{(0)} = 1$ and $B_2^{(0)} = 0$.

Then, since we perform our analysis in the Landau gauge, the gluon propagator, $\Delta_{\mu\nu}^{ab}(q)$, is strictly transverse,
\be 
\Delta^{ab}_{\mu\nu}(q) = - i\delta^{ab} P_{\mu\nu}(q) \Delta(q^2) \,, \qquad P_{\mu\nu}(q) := g_{\mu\nu} - \frac{q_\mu q_\nu}{q^2} \,.
\ee
%

Next, we take advantage of the Taylor theorem~\cite{Taylor:1971ff}, which states that in the Landau gauge the renormalization constant, $Z_1$, of the ghost-gluon vertex is finite. Furthermore, if we impose as renormalization condition that the vertex reduces to its tree level in the limit when the ghost momentum vanishes, \ie , $\Gnp_\mu(r,0,-r) = r_\mu$, then~\cite{Blossier:2010ky,Aguilar:2018csq} $Z_1 = 1$.

Requiring in addition that the propagators reduce to their tree levels, \ie \mbox{$F(\mu^2) = 1$} and $\Delta(\mu^2) = 1/\mu^2$, at an Euclidean momentum $\mu$ completely defines a self consistent renormalization scheme~\cite{Boucaud:2008ky,Boucaud:2008gn,Blossier:2010ky,Aguilar:2013xqa}, often called ``Taylor scheme''~\cite{Blossier:2010ky,Aguilar:2019jsj,Zafeiropoulos:2019flq}. For the present analysis, we fix the renormalization point at $\mu = 4.3$~GeV. 

Then, by virtue of the transversality of the gluon propagator and the choice of the Taylor scheme, one obtains from the ghost SDE of \fig{cpld}~\cite{Aguilar:2021okw}
\be 
F^{-1}(p^2) = 1 + \Sigma(p^2) - \Sigma(\mu^2) \,, \label{ghost_SDE}
\ee
where the ghost self-energy, $\Sigma(p^2)$, reads
\be
\Sigma(p^2) = ig^2 C_{\rm A} \int_k \! \Delta(k^2) D(s^2) f(k,p) B_1(-p, s, -k ) \,,
\ee
with $f(k,p) := 1 - (k\cdot p)^2/(k^2p^2)$, $s := k + p$, $C_{\rm A}$ the Casimir eigenvalue of the adjoint representation [$N$ for $SU(N)$], $g$ is the coupling constant, and we introduce the integral measure \mbox{$\int_k := \frac{1}{(2\pi)^4}\int d^4k$}. 

Note that, also due to the transversality of $\Delta^{ab}_{\mu\nu}(q)$, only the form factor $B_1$ of the ghost-gluon vertex contributes to the ghost SDE of \1eq{ghost_SDE}. This form factor can be extracted from the full vertex through the projector
\be 
\varepsilon_\mu(r,q) := \frac{q^2 r^\mu - q^\mu( q\cdot r) }{q^2 r^2 - (q\cdot r )^2 } \,. \label{eps_def}
\ee

Then, applying \1eq{eps_def} to the second line of \fig{cpld} yields a dynamical equation for the form factor $B_1(r,p,q)$, namely
\be
B_1(r,p,q) = 1 + \frac{ig^2C_{\rm A}r_\alpha p_\beta}{2} \left[ (d_1)^{\alpha\beta} - (d_2)^{\alpha\beta} \right] \,, \label{vertex_SDE}
\ee
with the $(d_i)^{\alpha\beta}$ denoting the contributions from the correspondingly named diagrams in \fig{cpld} and read~\cite{Aguilar:2021okw}
\begin{align}
(d_1)^{\alpha\beta} =& \int_k \! \Delta(k^2) \Delta(t^2) D(\ell^2) B_1( - \ell, p, t) \varepsilon^\mu(r,q) P^{\beta\rho}(t) P^{\alpha\sigma}(k)\Gnp_{\mu\sigma\rho}(q,k,-t) \,, \nonumber\\
(d_2)^{\alpha\beta} =& \int_k \! D(k^2) D(t^2) \Delta(\ell) B_1(k,-t,q)B_1(t,p,-\ell) \epsilon^\mu(q,k) k_\mu P^{\alpha\beta}(\ell) \,, \label{di} 
\end{align}
where $t := k + q$ and $\ell := k - r$.

For \2eqs{ghost_SDE}{vertex_SDE} to be a closed system of equations for $F(p^2)$ and $B_1(r,p,q)$, we must provide externally the gluon propagator and the three-gluon vertex which appears in the $(d_1)^{\alpha\beta}$ of \1eq{vertex_SDE} and is clearly the most complicated object. 

For the present work, as an approximation, we retain only the classical tensor structure of $\Gnp_{\alpha\mu\nu}(q,r,p)$, namely~\cite{Aguilar:2019jsj}
\begin{align}
\Gnp_{\alpha\mu\nu}(q,r,p) \approx &\,  (q - r )_\nu g_{\alpha\mu} X_1(q,r,p) + ( r - p )_\alpha g_{\mu\nu} X_1(r,p,q) + ( p - q )_\mu g_{\nu\alpha} X_1(p,q,r) \,, \label{3g_approx}
\end{align}
where the form factor $X_1(q,r,p)$ is symmetric under the exchange $q\leftrightarrow r$, such that \1eq{3g_approx} preserves the Bose symmetry of the vertex. At  tree level $X_1^{(0)}(q,r,p) = 1$.

The nonperturbative behavior of $X_1(q,r,p)$ can then be determined from the STI that the three-gluon vertex satisfies~\cite{Ball:1980ax,Aguilar:2019jsj}. Specifically,
\begin{align}
r^\mu \Gnp_{\alpha\mu\nu}(q,r,p) =&\, F(r^2) \left[ q^2 J(q^2) P^\mu_\alpha(q) H_{\mu\nu}(q,r,p) - p^2 J(p^2) P^\mu_\nu(p) H_{\mu\alpha}(p,r,q) \right] \,, \label{3g_STI}
\end{align} 
where $H_{\nu\mu}(q,r,p)$ is the ghost-gluon kernel~\cite{Aguilar:2018csq}, related to the ghost-gluon vertex by \mbox{$r^\nu H_{\nu\mu}(r,p,q) = \Gnp_\mu(r,p,q)$}, and $J(q^2)$ is the ``kinetic term'' of the gluon propagator, obtained by decomposing the latter as
\be 
\Delta^{-1}(q^2) = q^2 J(q^2) - m^2(q^2) \,,
\ee
where $m(q^2)$ is the dynamical gluon mass~\cite{Cornwall:1981zr,Aguilar:2006gr,Aguilar:2008xm,Aguilar:2019kxz,Aguilar:2020uqw}, which accounts for the IR saturation of $\Delta(q^2)$. Solving \1eq{3g_STI} allows us to express $X_1$ in terms of $F(p^2)$, $J(q^2)$ and certain form factors of $H_{\nu\mu}(q,r,p)$~\cite{Aguilar:2019jsj,Aguilar:2021okw}.

\begin{figure}[h]
  \includegraphics[width=0.5\linewidth]{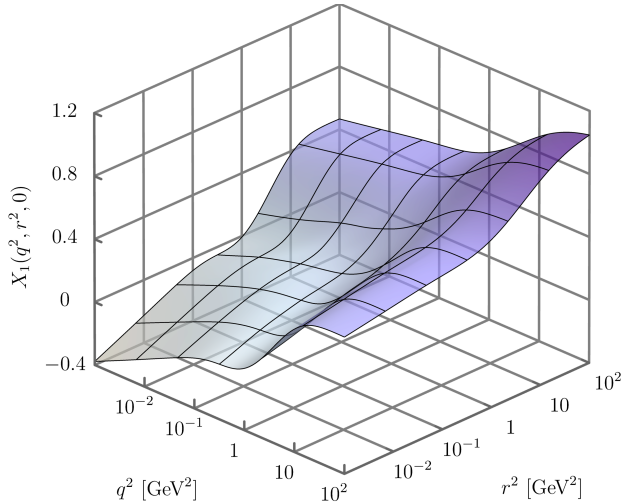}
  \caption{STI derived form factor $X_1(q,r,p)$ of the three-gluon vertex, for general $q^2$ and $r^2$, when the angle between $q$ and $r$ is $0$.\label{X1}}
\end{figure}

In \fig{X1} we show the result of this procedure for $X_1(q,r,p)$ for general $q^2$ and $r^2$, when the angle between $q$ and $r$ is zero~\cite{Aguilar:2021okw}, which is representative of its general kinematics behavior. In that figure, we see that although \1eq{X1} only retains $3$ out of the $14$ independent tensor structures~\cite{Ball:1980ax,Aguilar:2019jsj} of the three-gluon vertex, $X_1$ alone already encodes the most eminent features of this vertex, such as the positive anomalous dimension in the ultraviolet and the suppression with respect to the tree level in the IR~\cite{Huber:2012zj,Aguilar:2013vaa,Eichmann:2014xya,Blum:2015lsa,Athenodorou:2016oyh,Duarte:2016ieu,Boucaud:2017obn,Corell:2018yil,Aguilar:2019jsj,Aguilar:2019uob,Aguilar:2019kxz,Aguilar:2021lke,Catumba:2021yly}, driven by the masslessness of the ghosts~\cite{Aguilar:2013vaa}.

\begin{figure}[h]
  \includegraphics[width=0.5\linewidth]{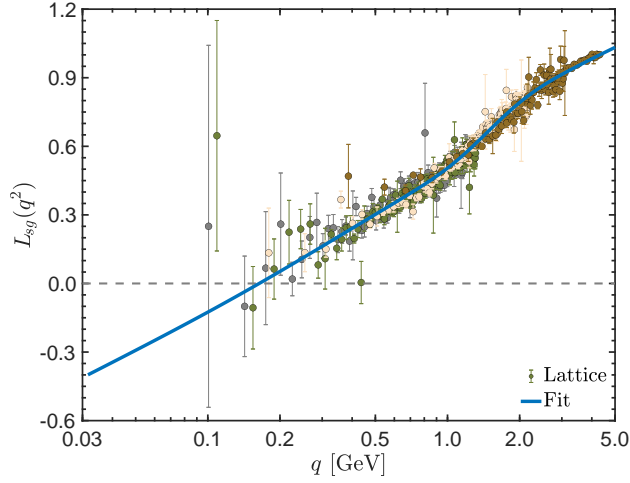}
  \caption{ Lattice results for the soft gluon projection, $\Ls(q^2)$, of the three-gluon vertex of Ref.~\cite{Aguilar:2021lke} (points) and the fit of Eq.~(4.6) of \cite{Aguilar:2021okw} (blue solid line). \label{Lasym}}
\end{figure}

An important consistency check on our approximations is provided by considering the soft gluon limit, $q = 0$, of the ghost-gluon vertex SDE. In this limit, \1eq{vertex_SDE} can be shown to reduce \emph{exactly} to~\cite{Aguilar:2021okw}
\begin{align}
B_1(r^2) =&\, 1 - \frac{ig^2 C_{\rm A} }{{\widetilde z}_3}\int_k D(\ell^2) \Delta^2(k^2) f(k,r) (r\cdot k) B_1(\ell,r,-k)\Ls(k^2) \nonumber\\
& + \frac{ig^2C_{\rm A}}{2}\int_k D^2(k^2)\Delta(\ell^2)f(k,r) \frac{k^2(r\cdot k)}{\ell^2} B_1(-k,r,\ell)B_1(k^2) \,, \label{sgl}
\end{align}
where $B_1(k^2):= B_1(k,-k,0)$ and $\Ls(q^2)$ is a special projection of the three-gluon vertex in the soft gluon limit, which has been accurately determined on the lattice~\cite{Aguilar:2021lke} and is shown in \fig{Lasym}. Lastly, ${\widetilde z}_3 \approx 0.95$ is a finite renormalization constant which converts $\Ls(q^2)$ from the ``asymmetric MOM" renormalization scheme employed on the lattice to the Taylor scheme used here~\cite{Aguilar:2021okw}.

Hence, while the right hand side of \1eq{sgl} still depends on the general kinematics $B_1$, for which we can use the result of the coupled system of \2eqs{ghost_SDE}{vertex_SDE}, its direct dependence on the three-gluon vertex is tightly controlled by employing the lattice results for $\Ls(q^2)$.

\begin{figure}[h]
  \includegraphics[width=0.5\linewidth]{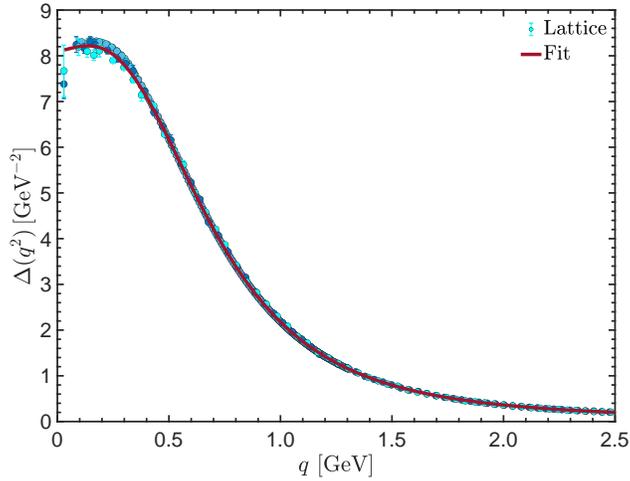}
  \caption{ Lattice data (circles) of \cite{Bogolubsky:2009dc,Boucaud:2018xup,Aguilar:2021okw} and a fit given by Eq.~(B5) of \cite{Aguilar:2021okw} (red solid line) for the gluon propagator. \label{gluon_fit}}
\end{figure}

Finally, we transform \3eqs{ghost_SDE}{vertex_SDE}{sgl} to Euclidean space, following \eg Eq.~(5.1) of~\cite{Aguilar:2018csq} and use for $\Delta(q^2)$ the fit of Eq.~(B5) of Ref.~\cite{Aguilar:2021okw} to lattice data of \cite{Bogolubsky:2009dc,Boucaud:2018xup,Aguilar:2021okw}, extrapolated to the continuum and to infinite volume, both shown in \fig{gluon_fit}.

\section{Results}

\begin{figure}[h]
  \includegraphics[width=0.5\linewidth]{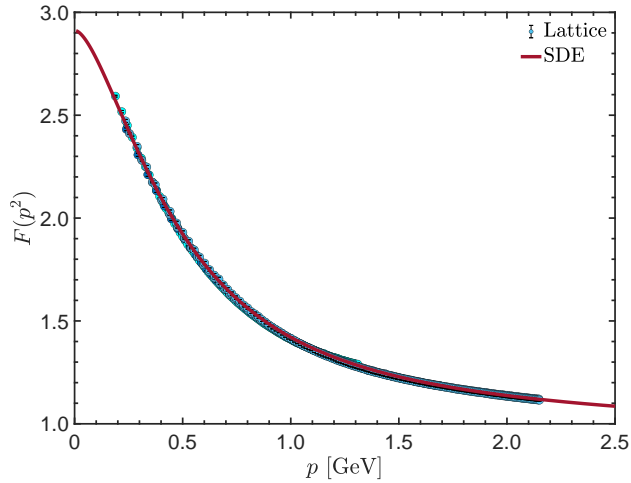}
  \caption{Ghost dressing function obtained by lattice simulations in Refs.~\cite{Boucaud:2018xup,Aguilar:2021okw} (points) compared to the solution of the coupled system of \2eqs{ghost_SDE}{vertex_SDE} (red solid line). \label{F_vs_lat}}
\end{figure}

We start by solving the coupled system composed of \2eqs{ghost_SDE}{vertex_SDE} for various values of $\alpha_s(\mu^2) = g^2/4\pi$, employing an iterative method. Then, we compare the resulting ghost dressing function, $F(p^2)$, to the lattice data of \cite{Boucaud:2018xup,Aguilar:2021okw}, which is cured from discretization artifacts. Minimizing the $\chi^2$, we obtain the value $\alpha_s(\mu^2) = 0.244$, and for $F(p^2)$ the red curve in \fig{F_vs_lat}, which agrees perfectly to the lattice result.

\begin{figure}[h]
  \includegraphics[width=0.5\linewidth]{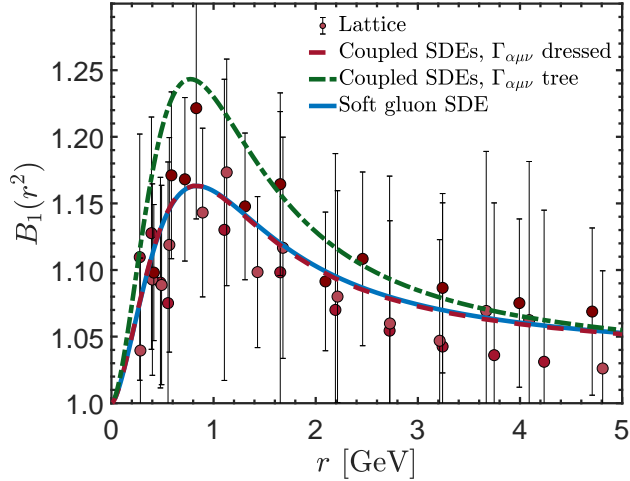}
  \caption{Soft gluon limit $B_1(r^2)$ of the ghost-gluon vertex obtained by lattice simulations in Ref.~\cite{Ilgenfritz:2006he} (circles), compared to the results of the coupled SDEs of \2eqs{ghost_SDE}{vertex_SDE} with the three-gluon vertex dressed (red dashed) or at tree level (green dot-dashed), and the result of the soft gluon SDE of \1eq{sgl} (blue continuous). \label{B1_soft}}
\end{figure}

Using the value of $\alpha_s(\mu^2)$ determined above, we compare in \fig{B1_soft} the soft gluon limit $B_1(r^2)$ obtained from the coupled system (red dashed line) to the lattice results of~\cite{Ilgenfritz:2006he} (circles), finding excellent agreement. Other kinematic limits are qualitatively similar to $B_1(r^2)$ and are shown in Ref.~\cite{Aguilar:2021okw}.

Next, to evaluate the impact of the three-gluon vertex in our results, we set it to tree level, which amounts to substituting $X_1\to 1$ in \1eq{3g_approx}. In this case, we obtain for the soft gluon $B_1(r^2)$ the green dot-dashed curve in \fig{B1_soft}. Comparing it to the red dashed line of the same figure we observe that the IR suppression of the nonperturbative three-gluon vertex has a significant impact on the radiative correction to $B_1(r^2)$, \ie in the quantity $B_1(r^2) - 1$, which is reduced by $30\%$ when $X_1$ is dressed.

The solution of the coupled system for $F(p^2)$ and the general kinematics $B_1(r,p,q)$ are then used as inputs in the soft gluon SDE of \1eq{sgl}, together with the fit for the lattice $\Ls(q^2)$ of Ref.~\cite{Aguilar:2021lke} shown in \fig{Lasym} and given by Eq.~(B5) of \cite{Aguilar:2021okw}. The solution $B_1(r^2)$ (blue continuous line in \fig{B1_soft}) is then compared to the result of the coupled system and found to coincide almost exactly, indicating the accuracy of the STI-derived approximation for the three-gluon vertex. 

\section{Conclusions}

We conducted a detailed study of the coupled dynamics of the ghost dressing function, $F(p^2)$, and the classical form factor of the ghost-gluon vertex, $B_1(r,p,q)$, through the SDEs that they satisfy. Within our truncation, the system of SDEs composed of \2eqs{ghost_SDE}{vertex_SDE} takes as inputs lattice data for the gluon propagator and an STI-derived approximation for the general kinematics three-gluon vertex, leaving a single parameter, namely the strong coupling constant, to be adjusted by matching $F(p^2)$ to lattice results.

For the value $\alpha_s(4.3 \text{ GeV}) = 0.244$, both $F(p^2)$ and $B_1(r,p,q)$ display nearly perfect agreement to lattice results, as can be seen in  Figs.~\ref{F_vs_lat} and \ref{B1_soft}, and qualitative agreement to many previous studies~\cite{Schleifenbaum:2004id,Huber:2012kd,Aguilar:2013xqa,Cyrol:2016tym,Mintz:2017qri,Aguilar:2018csq,Huber:2018ned,Aguilar:2019jsj,Barrios:2020ubx,Huber:2020keu}. We emphasize, however, that for $F(p^2)$ this level of agreement is only achieved when comparing to the lattice data of~\cite{Boucaud:2018xup,Aguilar:2021okw} which have been cured from scale setting and discretization artifacts, as explained in detail in \cite{Aguilar:2021okw}. This find explains a discrepancy found in previous works~\cite{Aguilar:2013xqa,Aguilar:2018csq}, between SDE results for $F(p^2)$ and the non extrapolated lattice data of \cite{Bogolubsky:2009dc}, stressing the importance of treating for lattice artifacts when comparing to predictions of continuum methods.

Our study controls for the effect of the three-gluon vertex by using as benchmark the soft gluon limit of the SDE for $B_1$, in which the contribution from the three-gluon vertex becomes \emph{exactly}~\cite{Aguilar:2021okw} the lattice determined $\Ls(q^2)$~\cite{Aguilar:2021lke}. Then we find, as shown in \fig{B1_soft}, that the IR suppression furnished by the three-gluon vertex has a significant quantitative effect, reducing $B_1-1$ by $30\%$.

Finally, the agreement shown in \fig{B1_soft} between the results for $B_1(r^2)$ obtained from the soft gluon SDE and the coupled system indicate that the contributions to the three-gluon vertex omitted in the STI-derived Ansatz of \1eq{3g_approx} are subleading, in accord with \cite{Aguilar:2021lke,Aguilar:2019jsj}.

\section{Acknowledgments}

The author thanks the organizers of the 19\textsuperscript{th} International Conference on Hadron Spectroscopy and Structure for the opportunity. The work of M. N. F. is supported by FAPESP, under the grant No. 2020/12795-1.

\end{document}